\documentclass[preprintnumbers,amsmath,amssymb,showpacs]{revtex4}

\usepackage{graphicx}
\usepackage{dcolumn}
\usepackage{bm}
\usepackage{subfigure}

\begin{document}

\title{ Detection of $k$-partite entanglement  and   $k$-nonseparability  of  multipartite quantum states}

\author{Yan Hong}
 \affiliation {School of Mathematics and Science, Hebei GEO University, Shijiazhuang 050031,  China}

\author{Ting Gao}
\email{gaoting@hebtu.edu.cn} \affiliation {School of Mathematical Sciences, Hebei Normal University, Shijiazhuang 050024,  China}

\author{Fengli Yan}
\email{flyan@hebtu.edu.cn} \affiliation {College of Physics, Hebei Normal University, Shijiazhuang 050024,  China}

\begin{abstract}

Identifying the  $k$-partite entanglement  and   $k$-nonseparability  of  general $N$-partite  quantum states are fundamental issues in quantum information theory.   By use of  computable inequalities of nonlinear operators, we present some simple and powerful $k$-partite entanglement  and  $k$-nonseparability  criteria  that works very well and allow for a simple and inexpensive test for the whole hierarchy of $k$-partite entanglement and $k$-separability of $N$-partite systems with $k$ running from $N$ down to 2. We illustrate their strengths by considering several examples in which our criteria perform better than other known detection criteria. We are able to detect $k$-partite entanglement  and $k$-nonseparabilty of multipartite systems which have previously not been identified.
In addition, our results can be implemented  in today's experiments.
\end{abstract}


\maketitle

\section{Introduction}
Quantum entanglement is an important physical resource in quantum  computation and quantum information processing  and is also a key feature that distinguishes quantum theory from classical theory. So, it is very valuable for studying  the characterization and detection of entanglement for general multipartite systems.

There are two different ways to characterize the entanglement of multipartite systems \cite{RPMK09,Guhne2009}, one is according to the question "How many partitions are separable?", the other is "How many particles are entangled". The former is described by $k$-separability, the latter by $k$-partite entanglement. $k$-separability provides a fine graduation of states according
to their degrees of separability, while $k$-partite entanglement provides the hierarchic classification of states  according to their degrees of entanglement. Detecting and characterizing $k$-partite entanglement  provide a refined insight in entanglement dynamics.

 For $N$-partite quantum systems,  the $k$-partite entanglement and $k$-nonseparability  are two different concepts of multipartite entanglement except  that $2$-partite entanglement is equivalent to $N$-nonseparability and genuine  $N$-partite entanglement is equivalent to $2$-nonseparability. Both $k$-nonseparability and
$k$-partite entanglement can be used to characterize multipartite entanglement. There are
some approaches to identify genuine $N$-partite entanglement such as based on spin-squeezing inequality \cite{VitaglianoHyllus11}  and elements of density matrix \cite{GuhneSeevinck10,HuberMintert10,GanHong10,GanHong11,WuKampermann12,ChenMa12},  Bell-type inequalities \cite{Seevinck02}, semidefinite program and state extensions \cite{Doherty02,Doherty04,Doherty05}, covariance matrices \cite{Gittsovich05}, etc.

The $k$-partite entanglement of $N$-partite quantum systems has been studied by using different tools and some progress has been made.
Some of the complete set of generalized spin squeezing inequalities  employed to detect $k$-particle
entanglement and bound entanglement \cite{VitaglianoHyllus11}. A hierarchic classification of all states from $2$-partite entanglement to $N$-partite entanglement based on Wigner-Yanase skew information was presented \cite{ChenZQ2011}. Entanglement criteria in terms of the quantum Fisher information \cite{Hyllus2012,Goth2012} were applied to detect several classes of $k$-partite entanglement. Further approach had led to $k$-particle entanglement criteria which was developed  by a comparison of the Fisher information and the sum of variances of all local operators \cite{Gessner16}.

In recent years, $k$-nonseparability of $N$-partite quantum states has attracted more and more attention and extensively explored  in various ways \cite{GaoYan14,GaoHong13,HongGaoYan12,ABMarcus10,Ananth15, HuberLlobet13, HongLuoSong15,HongLuo16,KlocklHuber15}. Some practical convenient inequalities only involving elements of density matrix for detecting $k$-nonseparability  were constructed in \cite{GaoYan14,GaoHong13,HongGaoYan12,ABMarcus10,Ananth15}, which can be used to  distinguish $N$-1 different classes of any high dimensional $N$-partite $k$-inseparable states.
In \cite{HuberLlobet13} a method that derived from some inequalities in the form of entropy vector was developed  to test $k$-nonseparability, whose essence is also elements of density matrix.
 $k$-nonseparable states  can also be checked by quantum Fisher information and local uncertainty relations \cite{HongLuoSong15,HongLuo16}.
From the Bloch vector decomposition of a density matrix, the tensor elements  can reflect many important characteristics of multipartite quantum states. The correlation tensor norms used for determining $k$-nonseparability in multipartite states \cite{KlocklHuber15}.

Much effort has been put into the detection and
characterization of $k$-nonseparability and $k$-partite entanglement in multipartite systems, but it
is far from perfect because of the extremely complex structure of entanglement of multipartite quantum states \cite{RPMK09,Guhne2009}.
This paper is devoted to giving powerful inequalities that provide, upon violation, experimentally accessible sufficient conditions for $k$-nonseparability and $k$-partite entanglement in $N$-partite systems. Compared with previous works, the resulting criteria are stronger.  The inequalities for $k$-nonseparability detection given in \cite{Ananth15} can be seen as special case of our results.

In this paper, we will  further study the structure of entanglement  of multipartite quantum states and present some inequalities to detect $k$-partite entanglement and $k$-nonseparability of  multipartite quantum states.
In Sec. II,  some concepts  and  symbols   are introduced.
In Sec. III and Sec. IV, we derive a full hierarchy of  $k$-partite entanglement criteria and $k$-nonseparability criteria based on  linear local operators to characterize multipartite entanglement.
Finally, we show their application ability   and point out that they can be adapted in present-day experiments.

\section{ Definitions }

The $k$-partite entanglement and  $k$-nonseparability are two different concepts involving the  partitions of subsystem in $N$-partite quantum system.

An $N$-partite quantum pure state  $|\phi\rangle$ is said to be  $k$-producible if it can be written as a tensor product $|\phi\rangle=\bigotimes\limits_{i=1}^n|\phi_{A_i}\rangle$
such that each $|\phi_{A_i}\rangle$ is a state of at most $k$ particles \cite{Guhne2009}.
A pure state contains  $k$-partite entanglement if it is not ($k$-1)-producible.
For an $N$-partite mixed state $\rho$,  if  it can be written as a convex combination of  $k$-producible pure states, then it is called $k$-producible.
The individual pure states composing a $k$-producible mixed state may be $k$-producible under different partitions.
If  $\rho$ is not $k$-producible, we say that $\rho$ contains $(k+1)$-partite entanglement. Here $1 \leq k\leq N-1$.

An $N$-partite quantum pure state $|\phi\rangle$  is called $k$-separable if there is a splitting  of $N$ particles into $k$ partitions $A_1,A_2,\cdots,A_k$  such that $|\phi\rangle=\bigotimes\limits_{i=1}^k|\phi_{A_i}\rangle$  \cite{Guhne2009}.
For an $N$-partite mixed state $\rho$,  if it can be written as a convex combination of  $k$-separable pure states, then it is called $k$-separable.
If  $\rho$ is not $k$-separable, we say that $\rho$ is $k$-nonseparable. Here $2 \leq k\leq N$.

In particular, the $1$-producible (or $N$-separable) states are  just fully separable states, and the quantum states being not $(N-1)$-producible (or 2-separable) are  genuinely entangled states.

For the following statement, let's  introduce some symbols.
Let $P$ be the global permutation operator performing simultaneous permutations on all subsystems in $(\mathcal{H}_1\otimes \mathcal{H}_2\otimes\cdots\otimes \mathcal{H}_N)^{\otimes2}$,
while $P_\alpha$ be local permutation operators  permuting the two copies of all subsystems contained in the set $\alpha$, that is,
$$P\Big(\bigotimes\limits_{i=1}^N|x_i\rangle\Big)\Big(\bigotimes\limits_{i=1}^N|y_i\rangle\Big)=\Big(\bigotimes\limits_{i=1}^N|y_i\rangle\Big)\Big(\bigotimes\limits_{i=1}^N|x_i\rangle\Big),$$
$$P_\alpha\Big(\bigotimes\limits_{i=1}^N|x_i\rangle\Big)\Big(\bigotimes\limits_{i=1}^N|y_i\rangle\Big)
=\Big(\bigotimes\limits_{i\in\alpha}|y_i\rangle\bigotimes\limits_{i\notin\alpha}|x_i\rangle\Big)\Big(\bigotimes\limits_{i\in\alpha}|x_i\rangle\bigotimes\limits_{i\notin\alpha}|y_i\rangle\Big).$$

\section{  Detection of  $k$-partite entanglement}

Now we state our main results in the detection of  $k$-partite entanglement.

 \emph{Theorem 1.} If $\rho$ is a $k$-producible quantum state of  $N$-partite quantum system $\mathcal{H}_1\otimes\mathcal{H}_2\otimes\cdots\otimes \mathcal{H}_N$,  where dim $\mathcal{H}_i=d_i, i=1,2,\cdots,N$, then
\begin{equation}\label{kproducible1}
\begin{array}{rl}
(2^{r}-2)\sqrt{\langle\Phi|\rho^{\otimes2}P|\Phi\rangle}\leq\sum\limits_{\{\alpha\}}\sqrt{\langle\Phi| P_\alpha^\dagger\rho^{\otimes2}P_\alpha|\Phi\rangle},
\end{array}
\end{equation}
where $r=\frac{N}{k}$ for $k|N$, $r=[\frac{N}{k}]+1$ for $k\nmid N$,  $|\Phi\rangle=|\phi_1\rangle|\phi_2\rangle$ with $|\phi_1\rangle=\bigotimes\limits_{i=1}^N|x_i\rangle$ and  $|\phi_2\rangle=\bigotimes\limits_{i=1}^N|y_i\rangle$ being any fully separable states of $N$-partite quantum system, and $\{\alpha\}$  consists of all possible nonempty proper subsets of $\{1,2,\cdots,N\}$. That is, an $N$-partite state $\rho$ does not satisfy inequality (\ref{kproducible1}), then $\rho$ contains  $k+1$-partite entanglement.

\emph{Proof.} Let us first  prove that inequality (\ref{kproducible1}) holds for any $k$-producible pure state. Suppose that pure state  $\rho=|\varphi\rangle\langle\varphi|$ with $|\varphi\rangle=\bigotimes\limits_{i=1}^n|\varphi_{A_i}\rangle$ is $k$-producible, where every vector $|\varphi_{A_i}\rangle$ involves at most $k$ parties. Then we have
\begin{equation}\label{kproduciblepure1}
\begin{array}{rl}
&\langle\Phi|\rho^{\otimes2}P|\Phi\rangle=|\langle\phi_1|\varphi\rangle\langle\varphi|\phi_2\rangle|^2\\
=&\Big|\Big\{\prod\limits_{i=1}^n\Big[(\bigotimes\limits_{j\in A_i}\langle x_j|)|\varphi_{A_i}\rangle\Big]\Big\}\Big\{\prod\limits_{i=1}^n\Big[\langle\varphi_{A_i}|(\bigotimes\limits_{j\in A_i}| y_j\rangle)\Big]\Big\}\Big|^2.
\end{array}
\end{equation}
Let $\alpha_{j_1,\cdots,j_m}=A_{j_1}\cup A_{j_2}\cup\cdots\cup A_{j_m}$, $|\varphi_{\alpha_{j_1,\cdots,j_m}}\rangle=\bigotimes\limits_{t=1}^m|\varphi_{A_{j_t}}\rangle$, and  $|\varphi_{\overline{\alpha_{j_1,\cdots,j_m}}}\rangle=\bigotimes\limits_{t=m+1}^n|\varphi_{A_{j_t}}\rangle$, we derive
\begin{equation}\label{kproduciblepure2}
\begin{array}{rl}
&\langle\Phi| P_{\alpha_{j_1,\cdots,j_m}}^\dagger\rho^{\otimes2}P_{\alpha_{j_1,\cdots,j_m}}|\Phi\rangle\\
=&\Big(\bigotimes\limits_{j\in{\alpha_{j_1,\cdots,j_m}}}\langle y_j|\bigotimes\limits_{j\notin{\alpha_{j_1,\cdots,j_m}}}\langle x_j|\Big)
\Big(\bigotimes\limits_{l=1}^n|\varphi_{A_l}\rangle\langle\varphi_{A_l}|\Big)
\Big(\bigotimes\limits_{j\in{\alpha_{j_1,\cdots,j_m}}}|y_j\rangle\bigotimes\limits_{j\notin{\alpha_{j_1,\cdots,j_m}}}|x_j\rangle\Big)\\
&\times\Big(\bigotimes\limits_{j\in{\alpha_{j_1,\cdots,j_m}}}\langle x_j|\bigotimes\limits_{j\notin{\alpha_{j_1,\cdots,j_m}}}\langle y_j|\Big)
\Big(\bigotimes\limits_{l=1}^n|\varphi_{A_l}\rangle\langle\varphi_{A_l}|\Big)
\Big(\bigotimes\limits_{j\in{\alpha_{j_1,\cdots,j_m}}}|x_j\rangle\bigotimes\limits_{j\notin{\alpha_{j_1,\cdots,j_m}}}|y_j\rangle\Big)\\
=&\Big|\Big\{\prod\limits_{i=1}^n\Big[(\bigotimes\limits_{j\in A_i}\langle x_j|)|\varphi_{A_i}\rangle\Big]\Big\}\Big\{\prod\limits_{i=1}^n\Big[\langle\varphi_{A_i}|(\bigotimes\limits_{j\in A_i}| y_j\rangle\Big]\Big\}\Big|^2.
\end{array}
\end{equation}
Combining (\ref{kproduciblepure1}) and (\ref{kproduciblepure2}) gives that
$$\begin{array}{ll}
\sqrt{\langle\Phi|\rho^{\otimes2}P|\Phi\rangle}=\sqrt{\langle\Phi| P_{\alpha_{j_1,\cdots,j_m}}^\dagger\rho^{\otimes2}P_{\alpha_{j_1,\cdots,j_m}}|\Phi\rangle},
\end{array}$$
which implies that inequality (\ref{kproducible1}) holds for any $k$-producible pure state.

Next, we prove inequality (\ref{kproducible1}) also holds for any $k$-producible mixed state. Suppose that $\rho=\sum\limits_ip_i\rho_i=\sum\limits_ip_i|\varphi_i\rangle\langle\varphi_i|$ is $k$-producible mixed state, where $\rho_i=|\varphi_i\rangle\langle\varphi_i|$ is $k$-producible. Then by triangle inequality and the Cauchy-Schwarz inequality, one has
$$\begin{array}{ll}
&(2^{r}-2)\sqrt{\langle\Phi|\rho^{\otimes2}P|\Phi\rangle}\\
\leq&(2^{r}-2)\sum\limits_ip_i\sqrt{\langle\Phi|\rho_i^{\otimes2}P|\Phi\rangle}\\
\leq&\sum\limits_ip_i\sum\limits_{\{\alpha\}}\sqrt{\langle\Phi| P_\alpha^\dagger\rho_i^{\otimes2}P_\alpha|\Phi\rangle}\\
=&\sum\limits_{\{\alpha\}}\sum\limits_i\sqrt{p_i\bigotimes\limits_{j\in\alpha}\langle y_j|\bigotimes\limits_{j\notin\alpha}\langle x_j|\rho_i\bigotimes\limits_{j\in\alpha}| y_j\rangle\bigotimes\limits_{j\notin\alpha}| x_j\rangle}
\sqrt{p_i\bigotimes\limits_{j\in\alpha}\langle x_j|\bigotimes\limits_{j\notin\alpha}\langle y_j|\rho_i\bigotimes\limits_{j\in\alpha}| x_j\rangle\bigotimes\limits_{j\notin\alpha}| y_j\rangle}\\
\leq &\sum\limits_{\{\alpha\}}\sqrt{\Big(\sum\limits_ip_i\bigotimes\limits_{j\in\alpha}\langle y_j|\bigotimes\limits_{j\notin\alpha}\langle x_j|\rho_i\bigotimes\limits_{j\in\alpha}| y_j\rangle\bigotimes\limits_{j\notin\alpha}| x_j\rangle\Big)
\Big(\sum\limits_ip_i\bigotimes\limits_{j\in\alpha}\langle x_j|\bigotimes\limits_{j\notin\alpha}\langle y_j|\rho_i\bigotimes\limits_{j\in\alpha}| x_j\rangle\bigotimes\limits_{j\notin\alpha}| y_j\rangle\Big)}\\
=&\sum\limits_{\{\alpha\}}\sqrt{\langle\Phi| P_\alpha^\dagger\rho^{\otimes2}P_\alpha|\Phi\rangle}.
\end{array}$$
The proof is complete.

\emph{Theorem 2.} Suppose that $\rho$ is an $N$-partite density
matrix acting on Hilbert space $\mathcal{H}^{\otimes N}$ with  dim $\mathcal{H}=d$.   Let $|\Psi^{st}_{ij}\rangle=|\psi^{s}_i\rangle|\psi^{t}_j\rangle$, where  $|\psi^{s}_i\rangle=|x_1\cdots x_{i-1}s x_{i+1}\cdots x_N\rangle$ and $|\psi^{t}_j\rangle=|x_1\cdots x_{j-1}t x_{j+1}\cdots x_N\rangle$ are fully separable states of $\mathcal{H}^{\otimes N}$. If $\rho$ is a $k$-producible,  then
\begin{equation}\label{kproducible2}
\begin{array}{rl}
&\sum\limits_{\substack{s,t\in\{\omega_1,\cdots,\omega_T\}\\1\leq i,j\leq N,i\neq j}}\sqrt{\langle\Psi^{st}_{ij}|\rho^{\otimes2}P|\Psi^{st}_{ij}\rangle}\\
\leq & \sum\limits_{\substack{s,t\in\{\omega_1,\cdots,\omega_T\}\\1\leq i,j\leq N,i\neq j}}\sqrt{\langle\Psi^{st}_{ij}|P_{i}^\dagger\rho^{\otimes2}P_{i}|\Psi^{st}_{ij}\rangle}
+T(k-1)\sum\limits_{\substack{s\in\{\omega_1,\cdots,\omega_T\}\\1\leq i\leq N}}\sqrt{\langle\Psi^{ss}_{ii}|P_{i}^\dagger\rho^{\otimes2}P_{i}|\Psi^{ss}_{ii}\rangle}
\end{array}
\end{equation}
for $2\leq k\leq N-1$, and
\begin{equation}\label{kproducible21}
\begin{array}{rl}
\sqrt{\langle\Psi^{st}_{ij}|\rho^{\otimes2}P|\Psi^{st}_{ij}\rangle}
\leq \sqrt{\langle\Psi^{st}_{ij}|P_{i}^\dagger\rho^{\otimes2}P_{i}|\Psi^{st}_{ij}\rangle}
\end{array}
\end{equation}
for $k=1$.
Here $\{|\omega_1\rangle,\cdots,|\omega_T\rangle\}\subseteq\mathcal{H}$ is a quantum state set.

Of course, if an $N$-partite state $\rho$ does not satisfy the above inequality (\ref{kproducible2}) (respectively, inequality (\ref{kproducible21})), then $\rho$ contains  $k+1$-partite entanglement ($2\leq k\leq N-1$) (respectively, 2-partite entanglement).

It should be pointed that there are $\frac{1}{2}n(n-1)$ inequalities in eq.(\ref{kproducible21})), and violation of any one of them implies 2-partite entanglement.

\emph{Proof.}  To establish the validity of ineq. (\ref{kproducible2}) and ineq. (\ref{kproducible21}) for all
$k$-producible states $\rho$, let us first verify that this is true
for any $k$-producible pure state $\rho$.

Suppose that pure state  $\rho=|\varphi\rangle\langle\varphi|$ is $k$-producible under partition $\{A_1,A_2,\cdots,A_n\}$, and $|\varphi\rangle=\bigotimes\limits_{l=1}^n|\varphi_{A_l}\rangle$. Then we have
\begin{equation}\label{kproducible3samepart}
\begin{array}{rl}
&\sqrt{\langle\Psi^{st}_{ij}|\rho^{\otimes2}P|\Psi^{st}_{ij}\rangle}
=\sqrt{\langle\psi^{s}_{i}|\rho|\psi^{s}_{i}\rangle\langle\psi^{t}_{j}|\rho|\psi^{t}_{j}\rangle}\\
\leq &\dfrac{\sqrt{\langle\Psi^{ss}_{ii}|P_{i}^\dagger\rho^{\otimes2}P_{i}|\Psi^{ss}_{ii}\rangle}+\sqrt{\langle\Psi^{tt}_{jj}|P_{j}^\dagger\rho^{\otimes2}P_{j}|\Psi^{tt}_{jj}\rangle}}{2}
\end{array}
\end{equation}
in case of $i, j$ in same part, and
\begin{equation}\label{kproducible3differentpart}
\begin{array}{rl}
&\sqrt{\langle\Psi^{st}_{ij}|\rho^{\otimes2}P|\Psi^{st}_{ij}\rangle}
=\sqrt{\langle\psi|\rho|\psi\rangle\langle\psi^{st}_{ij}|\rho|\psi^{st}_{ij}\rangle}
=\sqrt{\langle\Psi^{st}_{ij}|P_{i}^\dagger\rho^{\otimes2}P_{i}|\Psi^{st}_{ij}\rangle}
\end{array}
\end{equation}
in case of $i, j$ in different parts ($i \in A_l, j \in A_{l'}$ with
$l \neq l'$) where $|\psi^{st}_{ij}\rangle:=|x_1\cdots x_{i-1}sx_{i+1}\cdots x_{j-1}tx_{j+1}\cdots x_N\rangle$. Moreover, (\ref{kproducible3differentpart})  indicates that inequality (\ref{kproducible21}) hols with equality  for any 1-producible pure states.

Combining (\ref{kproducible3samepart}) and (\ref{kproducible3differentpart}) ptoduces
\begin{equation}\label{}
\begin{array}{rl}
&\sum\limits_{\substack{s,t\in\{\omega_1,\cdots,\omega_T\}\\1\leq i\neq j\leq N}}\sqrt{\langle\Psi^{st}_{ij}|\rho^{\otimes2}P|\Psi^{st}_{ij}\rangle}\\
=&\sum\limits_{\substack{s,t\in\{\omega_1,\cdots,\omega_T\}\\1\leq i,j\leq N\\i\in A_l, j\in A_{l'},l\neq l'}}\sqrt{\langle\Psi^{st}_{ij}|\rho^{\otimes2}P|\Psi^{st}_{ij}\rangle}
+\sum\limits_{\substack{s,t\in\{\omega_1,\cdots,\omega_T\}\\1\leq i,j\leq N\\i,j\in A_l, i\neq j}}\sqrt{\langle\Psi^{st}_{ij}|\rho^{\otimes2}P|\Psi^{st}_{ij}\rangle}\\
\leq&\sum\limits_{\substack{s,t\in\{\omega_1,\cdots,\omega_T\}\\1\leq i,j\leq N\\i\in A_l, j\in A_{l'},l\neq l'}}\sqrt{\langle\Psi^{st}_{ij}|P_{i}^\dagger\rho^{\otimes2}P_{i}|\Psi^{st}_{ij}\rangle}
+\sum\limits_{\substack{s,t\in\{\omega_1,\cdots,\omega_T\}\\1\leq i,j\leq N\\i,j\in A_l, i\neq j}}\dfrac{\sqrt{\langle\Psi^{ss}_{ii}|P_{i}^\dagger\rho^{\otimes2}P_{i}|\Psi^{ss}_{ii}\rangle}+\sqrt{\langle\Psi^{tt}_{jj}|P_{j}^\dagger\rho^{\otimes2}P_{j}|\Psi^{tt}_{jj}\rangle}}{2}\\
\leq&\sum\limits_{\substack{s,t\in\{\omega_1,\cdots,\omega_T\}\\1\leq i\neq j\leq N}}\sqrt{\langle\Psi^{st}_{ij}|P_{i}^\dagger\rho^{\otimes2}P_{i}|\Psi^{st}_{ij}\rangle}
+T(k-1)\sum\limits_{\substack{s\in\{\omega_1,\cdots,\omega_T\}\\1\leq i\leq N}}\sqrt{\langle\Psi^{ss}_{ii}|P_{i}^\dagger\rho^{\otimes2}P_{i}|\Psi^{ss}_{ii}\rangle}.
\end{array}
\end{equation}
Hence, inequality (\ref{kproducible2}) holds for any $k$-producible pure state.

Now, let us prove (\ref{kproducible2}) also holds for any $k$-producible mixed state. Suppose that  $\rho=\sum\limits_mp_m\rho_m=\sum\limits_mp_m|\varphi_m\rangle\langle\varphi_m|$ is $k$-producible mixed state, where $\rho_m=|\varphi_m\rangle\langle\varphi_m|$ is $k$-producible. Then by triangle inequality and the Cauchy-Schwarz inequality, one has
$$\begin{array}{ll}
&\sum\limits_{\substack{s,t\in\{\omega_1,\cdots,\omega_T\}\\1\leq i\neq j\leq N}}\sqrt{\langle\Psi^{st}_{ij}|\rho^{\otimes2}P|\Psi^{st}_{ij}\rangle}
\leq\sum\limits_mp_m\sum\limits_{\substack{s,t\in\{\omega_1,\cdots,\omega_T\}\\1\leq i\neq j\leq N}}\sqrt{\langle\Psi^{st}_{ij}|\rho_m^{\otimes2}P|\Psi^{st}_{ij}\rangle}\\
\leq&\sum\limits_mp_m\Big(  \sum\limits_{\substack{s,t\in\{\omega_1,\cdots,\omega_T\}\\1\leq i\neq j\leq N}}\sqrt{\langle\Psi^{st}_{ij}|P_{i}^\dagger\rho_m^{\otimes2}P_{i}|\Psi^{st}_{ij}\rangle}
+T(k-1)\sum\limits_{\substack{s\in\{\omega_1,\cdots,\omega_T\}\\1\leq i\leq N}}\sqrt{\langle\Psi^{ss}_{ii}|P_{i}^\dagger\rho_m^{\otimes2}P_{i}|\Psi^{ss}_{ii}\rangle} \Big)\\
\leq&\sum\limits_{\substack{s,t\in\{\omega_1,\cdots,\omega_T\}\\1\leq i\neq j\leq N}}\sum\limits_m\sqrt{(p_m\langle\psi|\rho_m|\psi\rangle)(p_m\langle\psi^{st}_{ij}|\rho_m|\psi^{st}_{ij}\rangle)}
+T(k-1)\sum\limits_{\substack{s\in\{\omega_1,\cdots,\omega_T\}\\1\leq i\leq N}}\sum\limits_mp_m\langle\psi^{s}_{i}|\rho_m|\psi^{s}_{i}\rangle\\
\leq&\sum\limits_{\substack{s,t\in\{\omega_1,\cdots,\omega_T\}\\1\leq i\neq j\leq N}}\sqrt{(\sum\limits_mp_m\langle\psi|\rho_m|\psi\rangle)(\sum\limits_mp_m\langle\psi^{st}_{ij}|\rho_m|\psi^{st}_{ij}\rangle)}
+T(k-1)\sum\limits_{\substack{s\in\{\omega_1,\cdots,\omega_T\}\\1\leq i\leq N}}\langle\psi^{s}_{i}|(\sum\limits_mp_m\rho_m)|\psi^{s}_{i}\rangle\\
\leq&\sum\limits_{\substack{s,t\in\{\omega_1,\cdots,\omega_T\}\\1\leq i\neq j\leq N}}\sqrt{\langle\Psi^{st}_{ij}|P_{i}^\dagger\rho^{\otimes2}P_{i}|\Psi^{st}_{ij}\rangle}
+T(k-1)\sum\limits_{\substack{s\in\{\omega_1,\cdots,\omega_T\}\\1\leq i\leq N}}\sqrt{\langle\Psi^{ss}_{ii}|P_{i}^\dagger\rho^{\otimes2}P_{i}|\Psi^{ss}_{ii}\rangle},
\end{array}$$
as desired.

Next, we prove that (\ref{kproducible21}) also holds for any 1-producible mixed state.
Suppose that  $\rho=\sum\limits_mp_m\rho_m=\sum\limits_mp_m|\varphi_m\rangle\langle\varphi_m|$ is $1$-producible mixed state, where $\rho_m=|\varphi_m\rangle\langle\varphi_m|$ is $1$-producible. Then by (\ref{kproducible3differentpart}), triangle inequality and the Cauchy-Schwarz inequality, one has
$$\begin{array}{ll}
&\sqrt{\langle\Psi^{st}_{ij}|\rho^{\otimes2}P|\Psi^{st}_{ij}\rangle}
\leq\sum\limits_mp_m\sqrt{\langle\Psi^{st}_{ij}|\rho_m^{\otimes2}P|\Psi^{st}_{ij}\rangle}
=\sum\limits_mp_m\sqrt{\langle\Psi^{st}_{ij}|P_{i}^\dagger\rho_m^{\otimes2}P_{i}|\Psi^{st}_{ij}\rangle}\\
\leq&\sqrt{(\sum\limits_mp_m\langle\psi|\rho_m|\psi\rangle)(\sum\limits_mp_m\langle\psi^{st}_{ij}|\rho_m|\psi^{st}_{ij}\rangle)}
=\sqrt{\langle\psi|\rho|\psi\rangle\langle\psi^{st}_{ij}|\rho|\psi^{st}_{ij}\rangle}
=\sqrt{\langle\Psi^{st}_{ij}|P_{i}^\dagger\rho^{\otimes2}P_{i}|\Psi^{st}_{ij}\rangle}.
\end{array}$$
Hence, inequality (\ref{kproducible21}) holds for any 1-producible states. This complete the proof.

\section{ Detection of $k$-nonseparability}

Here we present our main results in the detection  $k$-nonseparability.

\emph{Theorem 3.} If $\rho$ is a $k$-separable $N$-partite quantum state acting on Hilbert space $\mathcal{H}_1\otimes \mathcal{H}_2\otimes\cdots\otimes \mathcal{H}_N$,  where dim $\mathcal{H}_i=d_i, i=1,2,\cdots,N$, then
\begin{equation}\label{kseparable1}
\begin{array}{rl}
(2^{k}-2)\sqrt{\langle\Phi|\rho^{\otimes2}P|\Phi\rangle}\leq\sum\limits_{\{\alpha\}}\sqrt{\langle\Phi| P_\alpha^\dagger\rho^{\otimes2}P_\alpha|\Phi\rangle}.
\end{array}
\end{equation}
Here $|\Phi\rangle=|\phi_1\rangle|\phi_2\rangle$ with $|\phi_1\rangle=\bigotimes\limits_{j=1}^N|x_j\rangle$ and  $|\phi_2\rangle=\bigotimes\limits_{j=1}^N|y_j\rangle$ being any fully separable states of $N$-partite quantum system, and $\{\alpha\}$  consists of all possible nonempty proper subsets of $\{1,2,\cdots,N\}$.

Of course, $\rho$ is a $k$-nonseparable $N$-partite state if it violates
the above inequality (\ref{kseparable1}).

The proof is similar to that of Theorem 1.

Criterion 1 of Ref. \cite{Ananth15} is the special case of Theorem 3  when $|\Phi\rangle=|00\cdots0\rangle|(d_1-1)(d_2-1)\cdots(d_N-1)\rangle$.

\emph{Theorem 4.} If $\rho$ is a $k$-separable $N$-partite quantum state acting on Hilbert space $\mathcal{H}^{\otimes N}$,  where dim $\mathcal{H}=d$, then
\begin{equation}\label{kseparable2}
\begin{array}{rl}
&\sum\limits_{\substack{s,t\in\{\omega_1,\cdots,\omega_T\}\\1\leq i,j\leq N,i\neq j}}\sqrt{\langle\Psi^{st}_{ij}|\rho^{\otimes2}P|\Psi^{st}_{ij}\rangle}\\
\leq& \sum\limits_{\substack{s,t\in\{\omega_1,\cdots,\omega_T\}\\1\leq i,j\leq N,i\neq j}}\sqrt{\langle\Psi^{st}_{ij}|P_{i}^\dagger\rho^{\otimes2}P_{i}|\Psi^{st}_{ij}\rangle}
+T(N-k)\sum\limits_{\substack{s\in\{\omega_1,\cdots,\omega_T\}\\1\leq i\leq N}}\sqrt{\langle\Psi^{ss}_{ij}|P_{i}^\dagger\rho^{\otimes2}P_{i}|\Psi^{ss}_{ii}\rangle},
\end{array}
\end{equation}
for $2\leq k\leq N-1$. Inequality (\ref{kproducible21}) holds for any $N$-separable states.
Here $\{|\omega_1\rangle,\cdots,|\omega_T\rangle\}\subseteq\mathcal{H}$,
$|\Psi^{st}_{ij}\rangle=|\psi^{s}_i\rangle|\psi^{t}_j\rangle$ with $|\psi^{s}_i\rangle=|x_1\cdots x_{i-1}s x_{i+1}\cdots x_N\rangle$ and $|\psi^{t}_j\rangle=|x_1\cdots x_{j-1}t x_{j+1}\cdots x_N\rangle$.

 If an $N$-partite state $\rho$ does not satisfy the above inequality
(\ref{kseparable2}), then $\rho$ is not k-separable (k-nonseparable).

The proof is similar to that of Theorem 2.

We need to emphasize that inequality (\ref{kproducible21}) holds for any $N$-separable states  because $N$-separable states are the same as 1-producible states in $N$-partite quantum system.

Criterion 2 of Ref. \cite{Ananth15} is the special case of Theorem 4  when $|\psi\rangle=|00\cdots0\rangle$ and $\{\omega_1,\cdots,\omega_T\}=\{1,2,\cdots,d-1\}$.

\section{ Illustration}

In this section, we illustrate our main results with some explicit examples. We indeed detects $k$-partite entangled states and  $k$-nonseparable
mixed multipartite states which beyond all previously studied criteria.

For convenience of comparison, we first list the criteria which  can identify  $k$-partite entanglement and $k$-separability of some quantum states.

 For an $N$-qubit quantum state $\rho$, let quantum Fisher information $F(\rho, H)=\sum\limits_{l,l'}\dfrac{2(\lambda_l-\lambda_{l'})^2}{(\lambda_l+\lambda_{l'})}|\langle l|H|l'\rangle|^2$, where $H=\frac{1}{2}\sum\limits_{i=1}^N\sigma_z^{(i)}$ with $\sigma_z^{(i)}=|0\rangle\langle0|-|1\rangle\langle1|$  acting  on the $i$-th qubit.

 (I)  If  $$F(\rho, H)>sk^2+(N-sk)^2,$$
then $\rho$ is not $k$-producible and contains $(k+1)$-partite entanglement \cite{Hyllus2012}. Here $s=\lfloor\frac{N}{k}\rfloor$ is the largest integer smaller than or equal
to $\frac{N}{k}$.

(II)  If  $$F(\rho, H)>(N-k+1)^2+k-1,$$
then $\rho$ is $k$-nonseparable \cite{HongLuoSong15}.

Suppose that $\rho$ is an $N$ partite state acting on Hilbert space $\mathcal{H}^{\otimes N}$ with dim$\mathcal{H}=d$. Let  $\{g_m\}_{m=1}^{d^2-1}$ be the SU($d$) generators and $G_m=\sum\limits_{i=1}^Ng_m^{(i)}$, where $g_m^{(i)}$ means that $g_m$ acts on the $i$-th subsystem.

(III)   If an $N$ partite state $\rho$ acting on Hilbert space  $\mathcal{H}^{\otimes N}$ satisfies
$$\sum\limits_{i=1}^{d^2-1}V(\rho,G_m)<\left\{\begin{array}{ll} 2N(d-2), & \quad \textrm{ if $N$ is even},\\\\
2N(d-2)+2, & \quad \textrm{ if $N$ is odd},
\end{array}\right.$$
then $\rho$ is not $2$-producible and  contains 3-partite entanglement \cite{VitaglianoHyllus11}.
 Here $V(\rho,G_m)=\textrm{tr}[\rho(G_m)^2]-[\textrm{tr}\rho(G_m)]^2$.

(IV) If $\rho$   is a $k$-separable $N$-partite quantum state acting on Hilbert space $\mathcal{H}^{\otimes N}$,  where dim $\mathcal{H}=d$. Then \cite{Ananth15}
\begin{equation}\label{}\nonumber
\begin{array}{rl}
&\sum\limits_{\substack{1\leq i,j\leq N\\p,q=1,2,\cdots,d-1}}|\rho_{p\times d^{N-i}+1,q\times d^{N-j}+1}|\\
\leq&\sum\limits_{\substack{1\leq i,j\leq N\\p,q=1,2,\cdots,d-1}}\sqrt{\rho_{1,1}\rho_{p\times d^{N-i}+q\times d^{N-j}+1,p\times d^{N-i}+q\times d^{N-j}+1}}
+(N-k)\sum\limits_{\substack{1\leq i\leq N\\p=1,2,\cdots,d-1}}\rho_{p\times d^{N-i}+1,p\times d^{N-j}+1}.
\end{array}
\end{equation}

\emph{Example 1.}  Consider the $10$-qubit mixed states,
\begin{eqnarray*}
\rho(p,q)=p|G\rangle\langle G|+q|\widetilde{G}\rangle\langle \widetilde{G}|+\frac{1-p-q}{2^{10}}\textbf{1},
\end{eqnarray*}
where $|G\rangle=\frac{1}{\sqrt{2}}(|0\rangle^{\otimes10}+|1\rangle^{\otimes10}),
|\widetilde{G}\rangle=\frac{1}{\sqrt{2}}(|0\rangle^{\otimes10}-i|1\rangle^{\otimes10}).$

We choose $|\Phi\rangle=|0\rangle^{\otimes10}|1\rangle^{\otimes10}$ for our Theorem 1 and Theorem 3.

The parameter ranges of 10-qubit mixed states $\rho(p,q)$ containing  $(k+1)$-partite entanglement (with $k$ =3 and $k$ = 4) detected by Theorem 1 and (I) are illustrated in Fig. 1. We can see that the quantum states containing 4-partite entanglement  in the area enclosed by
red line $a$, the $p$ axis, blue line $a'$  and $q$ axis are detected only by Theorem 1.
Similarly, we find that the quantum states containing 5-partite entanglement  in the area enclosed by
orange line $b$, the $p$ axis,  green line $b'$ and $q$ axis are detected only by Theorem 1.

For 10-qubit mixed states $\rho(p,q)$, the parameter ranges for $k$-nonseparability (with $k$ =3 and $k$ = 4) detected by Theorem 3 and (II) are illustrated in Fig. 2. We can see that these $3$-nonseparable quantum states in the area enclosed by
purple line $c$,  $p$ axis, cyan line $c''$,  line $q=1-p$, cyan line $c'$ and $q$ axis  are detected only by Theorem 3. Similarly, we find that these $4$-nonseparable quantum states  in the area enclosed by
magenta line $d$,  $p$ axis,  brown line $d''$,  line $q=1-p$, brown line $d'$ and $q$ axis are detected only by Theorem 3.

\begin{figure}
\begin{center}
  \subfigure{
    \label{}
    \includegraphics[scale=0.6]{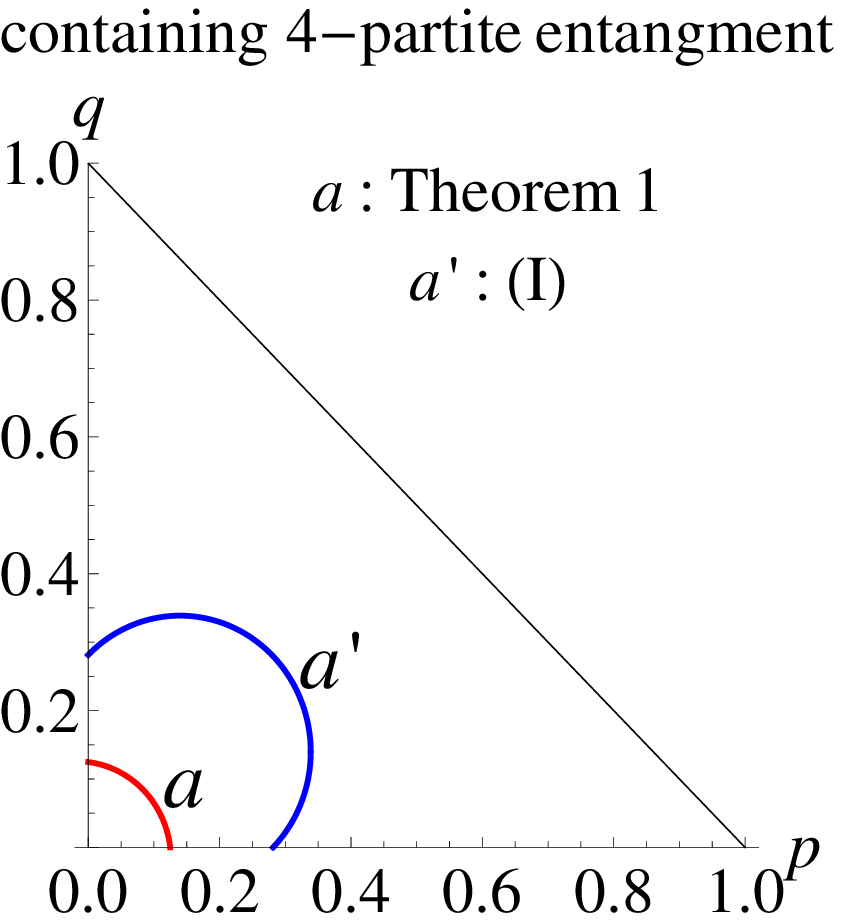}}
  \hspace{0.2in}
  \subfigure{
    \label{}
    \includegraphics[scale=0.6]{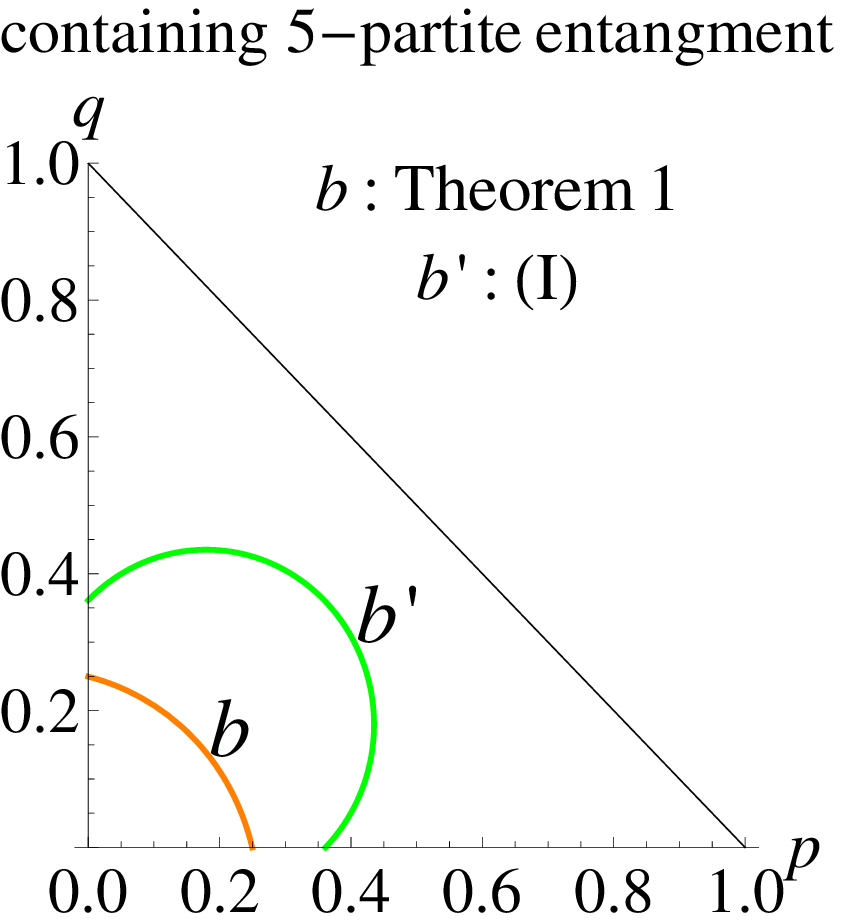}}
\caption[\label{}
]{  The 4-partite entanglement and 5-partite entanglement detection  for
$\rho(p,q)=p|G\rangle\langle G|+q|\widetilde{G}\rangle\langle \widetilde{G}|+\frac{1-p-q}{2^{10}}\textbf{1}$.
Here the red line $a$ represents the threshold of the detection for 4-partite entangled states given by Theorem 1, and
the area enclosed by
red line $a$, $p$ axis,  line $q=1-p$ and $q$ axis corresponds to the
quantum states containing 4-partite entanglement detected by Theorem 1;
 the blue line $a'$ represents the threshold given by (I),
the area enclosed
by blue line $a'$, $p$ axis,  line $q=1-p$ and $q$ axis corresponds to
the
quantum states containing 4-partite entanglement detected by (I). These quantum states containing 4-partite entanglement in the
area enclosed by red line $a$, $p$ axis, blue line $a'$ and $q$ axis are
detected only by  Theorem 1. Similarly, the orange line $b$ and green line $b'$ represent the
thresholds of the detection for 5-partite entangled states identified
by Theorem 1  and (I), respectively.
  these quantum states containing 5-partite entanglement in the
area enclosed by orange line $b$, $p$ axis, green line $b'$ and $q$ axis  are
detected only by  Theorem 1. }
  \end{center}
\end{figure}

\begin{figure}
\begin{center}
  \subfigure{
    \label{}
    \includegraphics[scale=0.6]{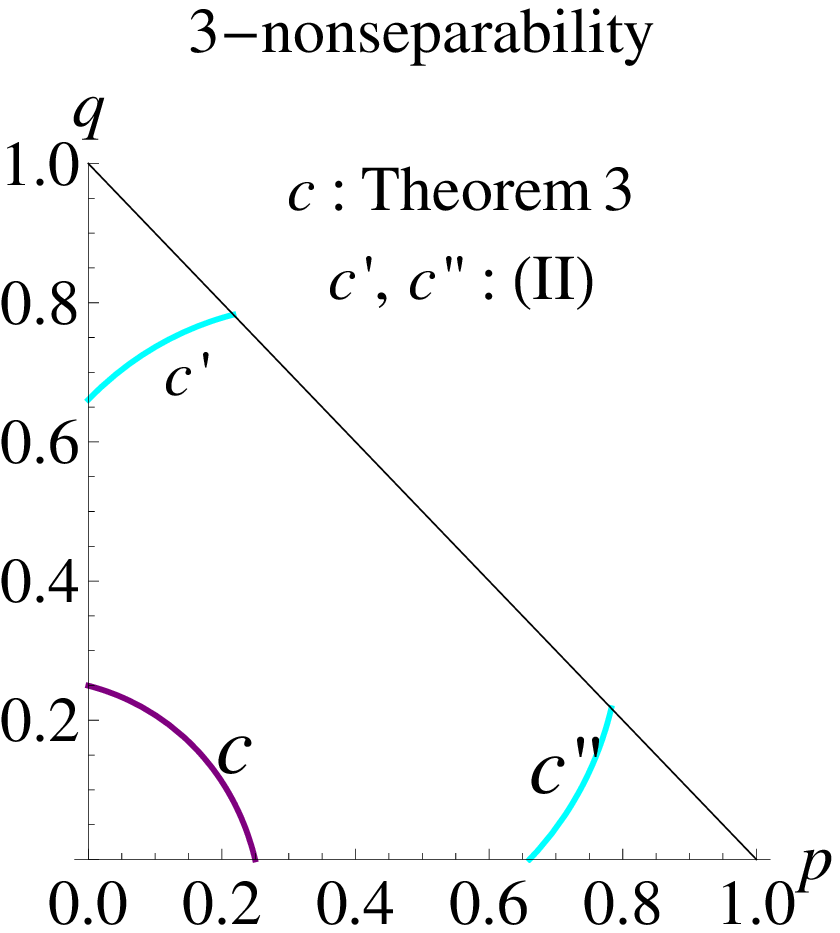}}
  \hspace{0.2in}
  \subfigure{
    \label{}
    \includegraphics[scale=0.6]{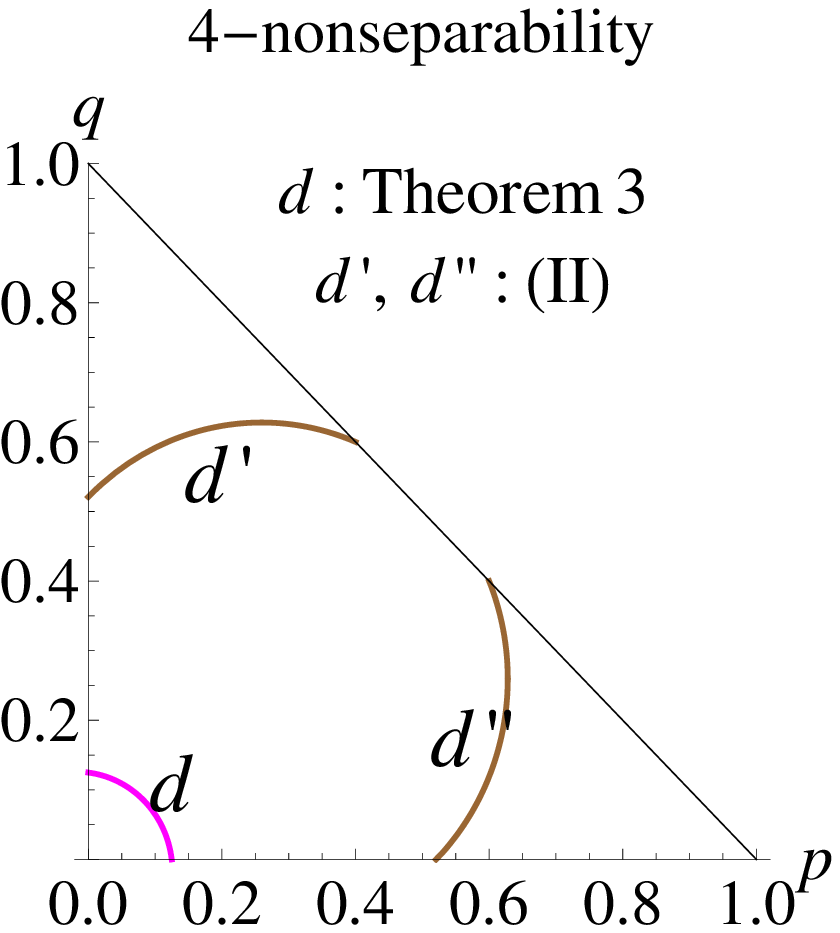}}
\caption[\label{}
]{  The detection power of Theorem 3 and (II) for
$\rho(p,q)=p|G\rangle\langle G|+q|\widetilde{G}\rangle\langle \widetilde{G}|+\frac{1-p-q}{2^{10}}\textbf{1}$
when $k=3$  and $k=4$.
Here the purple line $c$ represents the threshold given by Theorem 3, and
the area enclosed by
purple line $c$, $p$ axis,  line $q=1-p$ and $q$ axis corresponds to 3-nonseparable
quantum states detected by Theorem 3, while
 the cyan lines $c'$ and $c''$ represent the thresholds given by (II), and
the area enclosed
by cyan line $c''$, $p$ axis and line $q=1-p$ and the area enclosed
by cyan line $c'$, $q$ axis and line $q=1-p$ corresponds to
3-nonseparable
quantum states detected by (II). These 3-nonseparable quantum states in the
area enclosed by purple line $c$, $p$ axis, cyan line $c''$, line $q=1-p$, cyan line $c'$ and $q$ axis  are only
detected by  Theorem 3, but not by (II). Similarly, the magenta line $d$  represents the threshold of the detection for 4-nonseparable states identified by Theorem 3, the brown lines $c'$ and $c''$ represent the thresholds given by (II), and
  the
area encircled by  magenta line $d$, $p$ axis, brown line $d''$, line $q=1-p$, brown line $d'$ and $q$ axis  contains 4-nonseparable quantum states
detected only by  Theorem 3, but not by (II). }
  \end{center}
\end{figure}

\emph{Example 2.} Let us consider the family of 4-qutrit quantum states,
\begin{eqnarray*}
\rho(p,q)=p|W\rangle\langle W|+q\sigma^{\otimes4}|W\rangle\langle W|\sigma^{\otimes4}+\frac{1-p-q}{3^{4}}\textbf{1},
\end{eqnarray*}
where $|W\rangle=\frac{1}{2\sqrt{2}}(\sum\limits_{i=1}^{2}|i000\rangle+|0i00\rangle+|00i0\rangle+|000i\rangle),$ and $\sigma|0\rangle=|1\rangle$, $\sigma|1\rangle=|2\rangle$, $\sigma|2\rangle=|0\rangle$.

We choose $|\psi\rangle=|0\rangle^{\otimes4},\{\omega_1,\omega_2,\cdots,\omega_T\}=\{1,2\}$, and $|\psi\rangle=|1\rangle^{\otimes4}, \{\omega_1,\omega_2,\cdots,\omega_T\}=\{1,2\}$  for our Theorem 2 and Theorem 4 .

For 4-qutrit mixed states $\rho(p,q)$, Fig. 3 shows that Theorem 2 is more powerful than (III) for the detection of 3-partite entanglement.  These quantum states containing 3-partite entanglement in the
area enclosed by red line $e$, $p$ axis, blue line $f$ and $q$ axis only are
detected by  Theorem 2, but not by (III).

\begin{figure}
\begin{center}
    \label{}
    \includegraphics[scale=0.6]{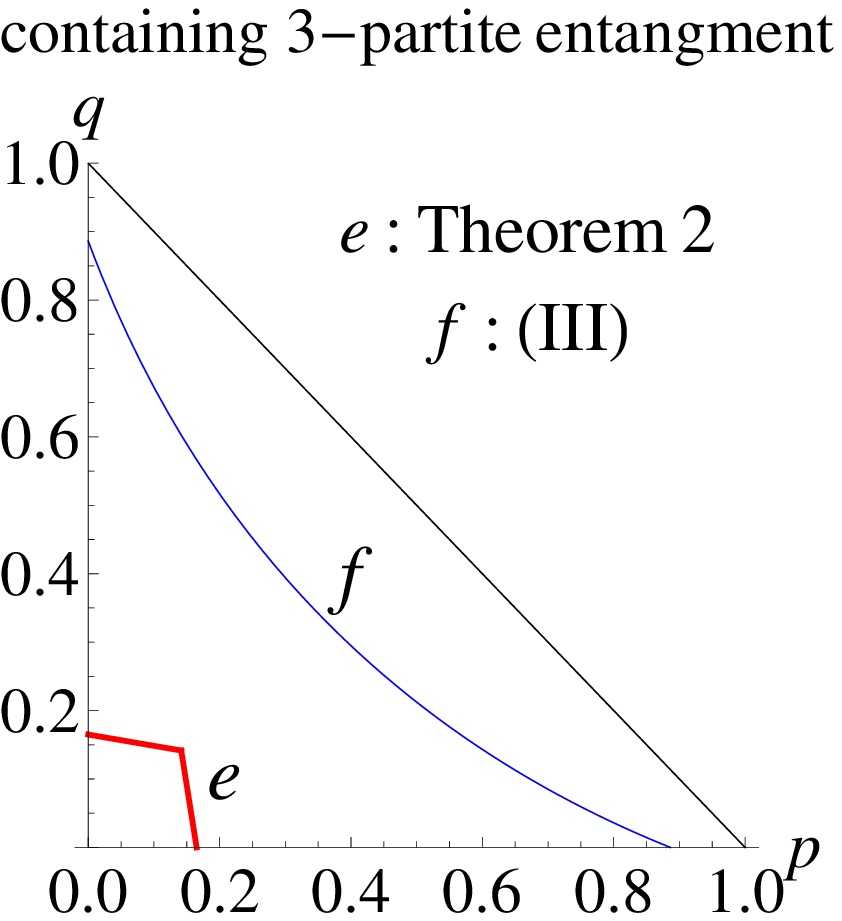}
\caption[\label{}
]{  The 3-partite entanglement detection  for
$\rho(p,q)=p|W\rangle\langle W|+q\sigma^{\otimes4}|W\rangle\langle W|\sigma^{\otimes4}+\frac{1-p-q}{3^{4}}\textbf{1}$.
The red line $e$ represents the threshold given by Theorem 2,  and
the area enclosed by
red line $e$, $p$ axis,  line $q=1-p$ and $q$ axis corresponds to the
quantum states containing 3-partite entanglement detected by Theorem 2.
The blue line $f$ represents the threshold given by  (III),
and
the area enclosed by
blue line $f$, $p$ axis,  line $q=1-p$ and $q$ axis corresponds to the
quantum states containing 3-partite entanglement detected by (II).
The  area enclosed by red line $e$, $p$ axis, blue line $f$ and $q$ axis are quantum states containing 3-partite entanglement
detected only by  Theorem 2, but not by (III). }
  \end{center}
\end{figure}

For these 4-qutrit mixed states $\rho(p,q)$, Theorem 4 is also stronger than (IV), which will be illuminated  in Fig. 4.
(IV)  can determine 3-nonseparability in the area enclosed by green line $g$, $p$ axis, and line $q=1-p$,
and Theorem 4 with $\{\omega_1,\omega_2,\cdots,\omega_T\}=\{1,2\}, |\psi\rangle=|0\rangle^{\otimes4}$ at the same time.
But Theorem 4 can identify other 3-nonseparable  quantum states  enclosed by orange line $h$,  line $q=1-p$ and $q$ axis with $\{\omega_1,\omega_2,\cdots,\omega_T\}=\{1,2\},,|\psi\rangle=|1\rangle^{\otimes4}$.
In other words, these 3-nonseparable states in the
area enclosed by green line $g$, orange line $h$, and $q$ axis  are
detected only by  Theorem 4, not by (IV).

\begin{figure}
\begin{center}
    \label{}
    \includegraphics[scale=0.6]{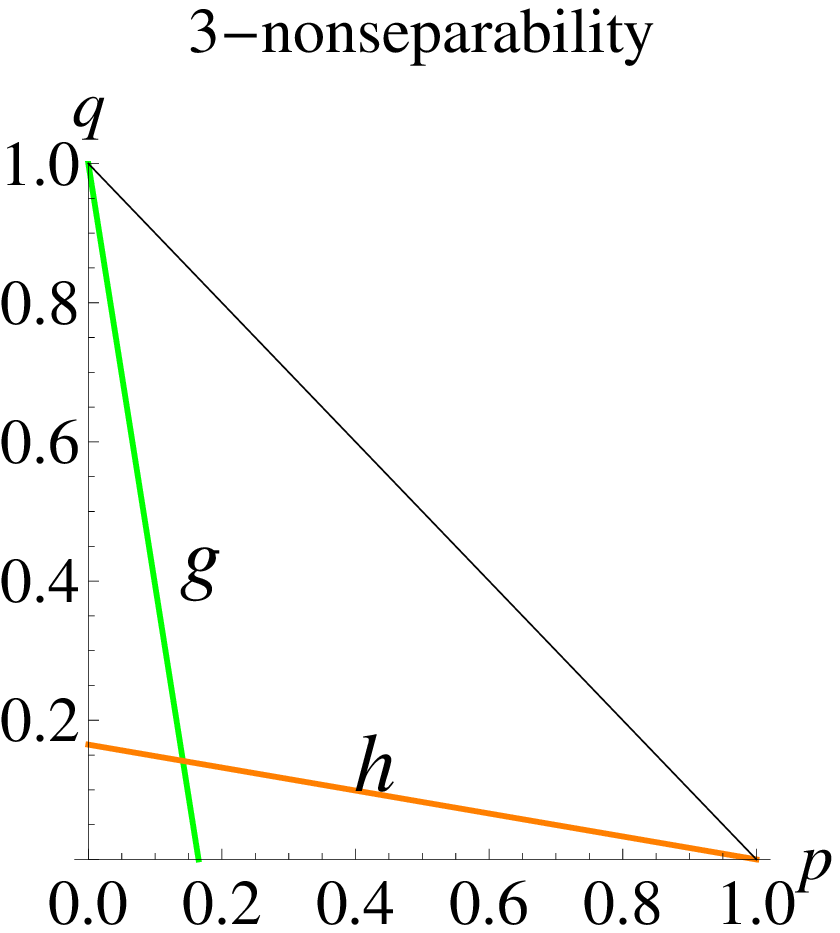}
\caption[\label{}
]{  The detection  of 3-nonseparability  for
$\rho(p,q)=p|W\rangle\langle W|+q\sigma^{\otimes4}|W\rangle\langle W|\sigma^{\otimes4}+\frac{1-p-q}{3^{4}}\textbf{1}$.
The green line $g$ represents the threshold given by Theorem 4 with $|\psi\rangle=|0\rangle^{\otimes4}$ and $\{\omega_1,\omega_2,\cdots,\omega_T\}=\{1,2\}$, and (IV) at the same time.
The orange line $h$ represents the threshold given by  Theorem 4  with $|\psi\rangle=|1\rangle^{\otimes4},$ and $\{\omega_1,\omega_2,\cdots,\omega_T\}=\{1,2\}$.
The area enclosed by
 green line $g$, $p$ axis, and line $q=1-p$  corresponds to the 3-nonseparable
 states  detected by Theorem 4 and (IV), the area enclosed by
orange line $h$,  line $q=1-p$ and $q$ axis corresponds to the 3-nonseparable
 states  detected by Theorem 4. Hence, the area enclosed by green line $g$, orange line $h$ and $q$ axis corresponds to the
3-nonseparable
 states  detected only by Theorem 4, not by (IV).

  }
  \end{center}
\end{figure}

\section{  experimental implementation}

Our  criteria can be implemented in today's experiment. Next we provide the local observables required to implement our criteria by using  the methods \cite{GanHong10,GaoHong13,GuhneLu2007,SeevinckUffink2008}.

The  left-hand sides of inequality (\ref{kproducible1}) and  (\ref{kseparable1}) $\sqrt{\langle\Phi|\rho^{\otimes2}P|\Phi\rangle}=|\langle\phi_1|\rho|\phi_2\rangle|$ can be implemented by two local observables $M$ and $\widetilde{M}$
because of $\langle M\rangle=2\textrm{Re}\langle\phi_1|\rho|\phi_2\rangle$ and $\langle \widetilde{M}\rangle=-2\textrm{Im}\langle\phi_1|\rho|\phi_2\rangle$,
where $M=|\phi_1\rangle\langle\phi_2|+|\phi_2\rangle\langle\phi_1|$ and $\widetilde{M}=-\textrm{i}|\phi_1\rangle\langle\phi_2|+\textrm{i}|\phi_2\rangle\langle\phi_1|$.
Let $|\phi_1\rangle=\bigotimes\limits_{n=1}^N|x_n\rangle$, $|\phi_2\rangle=\bigotimes\limits_{n=1}^N|y_n\rangle$, and
\begin{equation}\label{}\nonumber
\begin{array}{rl}
\mathcal{M}_l=&\bigotimes\limits_{n=1}^N\big[\cos(\frac{l\pi}{N})(|y_n\rangle\langle x_n|+|x_n\rangle\langle y_n|)+\sin(\frac{l\pi}{N})(\textrm{i}|y_n\rangle\langle x_n|-\textrm{i}|x_n\rangle\langle y_n|)\big],\\
\mathcal{\widetilde{M}}_l=&\bigotimes\limits_{n=1}^N\big[\cos(\frac{l\pi+\pi/2}{N})(|y_n\rangle\langle x_n|+|x_n\rangle\langle y_n|)+\sin(\frac{l\pi+\pi/2}{N})(\textrm{i}|y_n\rangle\langle x_n|-\textrm{i}|x_n\rangle\langle y_n|)\big],
\end{array}
\end{equation}
where $1\leq l\leq N$. Then they satisfy $\sum\limits_{l=1}^N(-1)^l\mathcal{M}_l=NM$, and $\sum\limits_{l=1}^N(-1)^l\mathcal{\widetilde{M}}_l=N\mathcal{\widetilde{M}}$.
The  right-hand side of inequality (\ref{kproducible1}) and  (\ref{kseparable1})  can be implemented by  local observables
$(\bigotimes\limits_{n\in\alpha}|y_n\rangle\langle y_n|)\otimes(\bigotimes\limits_{n\notin\alpha}|x_n\rangle\langle x_n|)$.

The  left-hand sides of  inequality (\ref{kproducible2}) and (\ref{kseparable2}) $\sqrt{\langle\Psi^{st}_{ij}|\rho^{\otimes2}P|\Psi^{st}_{ij}\rangle}=|\langle\psi_{i}^s|\rho|\psi_{j}^t\rangle|$ can be implemented by two local observables $M_{ij}^{st}$ and $\widetilde{M}_{ij}^{st}$
because of $\langle M_{ij}^{st}\rangle=4\textrm{Re}\langle\psi_{i}^s|\rho|\psi_{j}^t\rangle$ and $\langle \widetilde{M}_{ij}^{st}\rangle=-4\textrm{Im}\langle\psi_{i}^s|\rho|\psi_{j}^t\rangle$,
where
\begin{equation}\label{}\nonumber
\begin{array}{rl}
M_{ij}^{st}=&|\psi_{i}^s\rangle\langle\psi_{j}^t|+|\psi_{j}^t\rangle\langle\psi_{i}^s|\\
=&M_{i}^s\otimes M_{j}^t\otimes(\bigotimes\limits_{n\neq i,j}|x_n\rangle\langle x_n|)
+\widetilde{M}_{i}^s\otimes \widetilde{M}_{j}^t\otimes(\bigotimes\limits_{n\neq i,j}|x_n\rangle\langle x_n|),\\
\widetilde{M}_{ij}^{st}=&|\psi_{i}^s\rangle\langle\psi_{j}^t|+|\psi_{j}^t\rangle\langle\psi_{i}^s|\\
=&M_{i}^s\otimes \widetilde{M}_{j}^t\otimes(\bigotimes\limits_{n\neq i,j}|x_n\rangle\langle x_n|)
-\widetilde{M}_{i}^s\otimes M_{j}^t\otimes(\bigotimes\limits_{n\neq i,j}|x_n\rangle\langle x_n|).
\end{array}
\end{equation}
Here $M_{i}^s=|s\rangle\langle x_{i}|+|x_{i}\rangle\langle s|$,
$\widetilde{M}_{i}^s=\textrm{i}|s\rangle\langle x_{i}|-\textrm{i}|x_{i}\rangle\langle s|$.
Hence, The  left-hand sides of inequality (\ref{kproducible2}) and  (\ref{kseparable2}) can be implemented by local observables.
The  right-hand sides of inequality (\ref{kproducible2}) and  (\ref{kseparable2}) can be implemented by the local observables $\bigotimes\limits_{n=1}^N|x_n\rangle\langle x_n|, (|s\rangle\langle s|)\otimes(|t\rangle\langle t|)\otimes(\bigotimes\limits_{n\neq i,j}|x_n\rangle\langle x_n|)$ and $(|s\rangle\langle s|)\otimes(\bigotimes\limits_{n\neq i}|x_n\rangle\langle x_n|)$.

\section{Conclusions}

In conclusion, we exploit some nonlinear operators to develop a series of inequalities, which   can be used to efficiently identify  $k$-partite entanglement  and  $k$-nonseparability of $N$-partite mixed quantum states in arbitrary dimensional systems. These inequalities present sufficient conditions for  the detection of $k$-partite entanglement  and  $k$-nonseparability,  distinguish different classes of multipartite inseparable
states,  and can identify some $k$-partite entanglement  and  $k$-nonseparability that had not been identified so far.
These reveal the  practicability and efficiency of our results. Moreover,  our criteria can be applied to experiment and we give the corresponding local observables required to implement them.

\begin{center}
{\bf ACKNOWLEDGMENTS}
\end{center}

     This work was supported by  the National Natural Science
Foundation of China under Grant Nos. 12071110 and 11701135, the Hebei Natural Science
Foundation of China under Grant No. A2020205014 and No. A2017403025, the education Department of Hebei Province Natural Science Foundation under Grant No.  ZD2020167, and
the Foundation of  Hebei GEO University under Grant No. BQ201615.

\end{document}